# Proton-mediated reversible switching of metastable ferroelectric phases with low operation voltages

Xin He[1,2], Yinchang Ma[2], Chenhui Zhang[2], Aiping Fu[3], Weijin Hu[4,5], Yang Xu[1], Bin Yu[1], Kai Liu[6], Hua Wang[1]*, Xixiang Zhang[2]*, and Fei Xue[1,2]*

[1]ZJU-Hangzhou Global Scientific and Technological Innovation Center, School of Micro-Nano Electronics, Zhejiang University, Hangzhou 310020, China.
[2]Physical Science and Engineering Division, King Abdullah University of Science and Technology, Thuwal 23955-6900, Saudi Arabia.
[3]College of Chemistry and Chemical Engineering, Qingdao University, Qingdao, 266071, China.
[4]Shenyang National Laboratory for Materials Science, Institute of Metal Research, Chinese Academy of Sciences, Shenyang 110016, China.
[5]School of Materials Science and Engineering, University of Science and Technology of China, Shenyang 110016, China.
[6]Physics Department, Georgetown University, Washington, DC 20057, USA.

*Email: xuef@zju.edu.cn;  daodaohw@zju.edu.cn;  xixiang.zhang@kaust.edu.sa

The exploration of ferroelectric phase transitions enables an in-depth understanding of ferroelectric switching and promising applications in information storage. However, controllably tuning the dynamics of ferroelectric phase transitions remains challenging owing to inaccessible hidden phases. Here, using protonic gating technology, we create a series of metastable ferroelectric phases and demonstrate their reversible transitions in van der Waals ferroelectric $α$-$In_2Se_3$ transistors. By varying the gate bias, protons can be incrementally injected or extracted, achieving controllable tuning of the ferroelectric $α$-$In_2Se_3$ protonic dynamics across the channel and obtaining numerous intermediate phases. Combining piezoelectric force microscopy and electrical hysteresis measurements, we surprisingly discover that the gate tuning of $α$-$In_2Se_3$ protonation is volatile, unlike the conventional gating effect, and the created new phases remain polar. Their origin, revealed by first-principles calculations, is related to the formation of metastable hydrogen-stabilized $α$-$In_2Se_3$ phases with switchable dipoles. Furthermore, our approach enables ultralow gate-voltage switching of different phases (below 0.4 V), and easy integration with transistor components. This work provides fundamental insights into the interactions between protons and ferroelectric phase transitions, providing a possible avenue for accessing hidden phases in ferroelectric switching and for developing energy-efficient, multilevel-switching memory devices.

Ferroelectric phase transitions exhibit intriguing physics including changes in the structural symmetry or polar ordering, which offers promising applications in information storage. To date, many approaches have been documented for enabling ferroelectric phase transitions to explore polar physics and device implications (*1-9*). For example, when the temperature exceeds an important value (i.e., the Curie point), ferroelectric-to-paraelectric transitions can occur; furthermore, the material concurrently undergoes structural evolution from non-centrosymmetric crystals with off-centered ions to centrosymmetric counterparts (*10-12*). Similarly, modifying the film thicknesses can induce phase/structural transitions because of the competition between polarization and depolarization fields: thicker films retain ferroelectric phases while thinner films easily become paraelectric (*13, 14*). Moreover, the application of an electric field is another common method for triggering ferroelectric phase transitions, by which parallel (i.e., ferroelectric) and antiparallel (i.e., antiferroelectric) polar ordering can be reversibly switched (*2, 5, 15, 16*). Despite multiple attempts to explore ferroelectric phase transitions, their "hidden" or intermediate phases, which contain rich physics, are still inaccessible on equilibrium phase diagrams.

Among various approaches to incite ferroelectric phase transitions, only the application of an electric field provides promise for developing novel memory devices with high performances (e.g., $10^{12}$ endurance and 10-ns operation speed) (*17-19*). However, these devices generally have only two digital resistance states owing to the lack of polar configurations beyond parallel/antiparallel ordering. In principle, a greater number of polar configurations offers broader possibilities for implementing multilevel, analog memory devices toward data-centric computing use (*20-23*). Accordingly, a novel strategy that can yield ferroelectric phase transitions with accessible, multiple polar phases (or hidden phases) is both fundamentally interesting and technologically important.

Using protonic gate technology as a new approach, we demonstrate reversible ferroelectric phase transitions among multiple metastable protonic $α$-In$_2$Se$_3$ phases in a transistor architecture (Note S1 for the discussion over the van der Waals ferroelectric $α$-In$_2$Se$_3$). Gate control over proton injection/extraction can render ferroelectric phase transitions between proton-sufficient and proton-deficient compounds (i.e., $α$-In$_2$Se$_3$H$_y$, $α$-In$_2$Se$_3$H$_x$, and $α$-In$_2$Se$_3$H$_i$; y>x>i). Piezoelectric force microscopy (PFM) mappings demonstrate that the proton-mediated phase transition indeed comprises a multitude of intermediate phases, arising from different protonation levels. Along with additional electrical measurements and first-principles calculations, we find that these new phases remain metastable and polar, which results in our observed volatile gate-tuning effect and hysteresis in drain-source curves. Moreover, our approach can enable ultralow gate-voltage (< 0.4 V) switching of ferroelectric phases, thus opening a new avenue for developing low-power multilevel switching memory devices.

**Device Structure, Calculations, and Mechanisms**

As previously reported (*24, 25*), we implemented protonic gate technology in a three-terminal transistor architecture (Fig. 1A). This technology is used to manipulate the electronic and magnetic properties of certain materials (*25-30*), because protons (H$^+$) are the smallest ionic defects that can easily diffuse into rigid materials and achieve the desired tuning. The penetration of protons into channel materials has been confirmed by various measurement methods (*24, 26*), thus proving the effectiveness of this strategy. In transistors, porous silica serves as a gate dielectric and essentially provides protons via the formation of Si-OH across the films or via the electrolysis of absorbed water into $H^+$ within silica (see Note S2 and fig. S1 for more details) (*24*). Before coating 190-nm-thick silica films, we deposited high-quality $Al_2O_3$ insulator films (10 nm) on bottom Pt gate electrodes to prevent the breakdown of the transistor dielectric.

Exfoliated ferroelectric $α$-$In_2Se_3$ multilayers (typically ~ 20-60 nm) were transferred onto porous silica/$Al_2O_3$/Pt heterostructures. Gate control biases are applied using Pt electrodes and can accelerate proton diffusion, whereas the bias polarity determines proton injection or extraction (Note S3). Because of room-temperature thermal activation (*31, 32*), certain protons within the silica films can naturally diffuse into ferroelectric $α$-$In_2Se_3$, making pure initial phase protonic ($α$-$In_2Se_3$+$xH^+$+$xe^-$ ↔ $α$-$In_2Se_3H_x$), particularly at the bottom portion. Despite the top portion remaining less protonated, a hybrid compound (i.e., $α$-$In_2Se_3H_x$) can form with a proton concentration gradient perpendicular to its surface. To clarify the structure and stability of this compound, we employed first-principles calculations and Raman spectroscopy (Fig. 1B-1D).

In theoretical calculations, we considered the protonation effect in multilayered samples (e.g., bilayer). The energy barrier for injecting hydrogen atoms into the $α$-$In_2Se_3$ phase is ~200 meV, predicted by the minimum energy pathways of the protonation from climbing image nudged elastic band calculations (fig. S2). This barrier can be easily overcome by applying an electric field, and can thus facilitate channel protonation, as described here. As shown in Fig. 1B, a single hydrogen (H) atom is placed in the bottom intralayer of a bilayer α-$In_2Se_3$ unit cell, other than the top intralayer. The most energetically favorable structure is that H atoms reside near the middle Se atom layer and only chemically bond with this Se layer because there is charge transfer from the Se atom to the H atom (fig. S3). Simultaneously, intralayer sliding between central Se and H atoms occurs due to the small, generalized stacking fault energy for entirely shifting In-Se covalent bonds. After structural relaxation, the protonation at the bottom quintuple-atom-layer shows no obvious impact on the top quintuple-atom-layers (i.e., no proximate effect) because of the weak van der Waals interactions. The crystal structure can relax to a metastable state: once the H atom is removed, the unit cell reverses to the initial $α$-$In_2Se_3$ structure, while, with the presence of the H atom, it remains stable. Surprisingly, the metastable phase exhibits non-centrosymmetric properties and holds both in-plane and out-of-plane dipoles. This signifies the existence of ferroelectricity, which originates from the structural asymmetry between the Se-H interaction and two adjacent In atoms.

Considering hydrogen migration from the bottom to the top, the scenarios of H atoms located at the van der Waals gaps are also theoretically examined (Fig. 1C and fig. S4). The calculation demonstrates that hydrogens are more energetically favorable to reside within the gaps. Remarkably, the protonated structures partially resemble β-$In_2Se_3$, but display in-plane and out-of-plane dipoles arising from the protonic Se atoms. Based on these calculated results, we conclude that, for a multilayer sample (> 2 layers), the lower quintuple-atom-layers can be fully protonic but the upper layers may remain intact, resulting in new intermediate phases. Therefore, the zero-proximity effect of protonation on neighboring layers ensures that numerous ferroelectric intermediate states ($α$-In$_2$Se$_3$H) can be created by slightly adjusting proton concentrations and protonic thicknesses. Note that these multiple phases play a crucial role in the implementation of multilevel-resistance memory devices toward brain-inspired computing.

After the theoretical discussion over protonic structures, we move to the Raman characterization of protonic $α$-In$_2$Se$_3$ phases. Upon applying 0 and +2 V gate biases, the compound is subjected to different levels of protonation in which higher voltages impart a higher degree of protonation. As shown in Fig. 1D, we observe a blue shift for two prominent Raman peaks located at around 90 cm$^{-1}$ (i.e., E mode) and 198 cm$^{-1}$ (i.e., A$_1$ mode), whereas no shift is discerned for the other two peaks at 104 cm$^{-1}$ and 180 cm$^{-1}$, respectively. The shifted Raman modes may be originated from specific interlayer shearing and breathing vibrations, which are enhanced by proton injection across the van der Waals gaps. The intralayer breathing modes corresponding to unshifted peaks are much less sensitive to the proton injection as shown in fig. S5. It is worth mentioning that the detected shift ($> 0.5$ cm$^{-1}$), though small, can be reliably captured by our Raman measurement with a resolution of 0.5 cm$^{-1}$. These results provide strong direct evidence for protonation across the channel.

**Multiple protonic phases**

Schematics of ferroelectric protonation levels, with different gate biases, are presented in Fig. 2A. When considering a gate bias of 0 V, a few protons can spontaneously intercalate into multilayered channels, forming an $α$-In$_2$Se$_3$H$_{x1}$ compound. Following previous works (*24, 26*), the bottom portion is inevitably protonated, while the top portion is only weakly protonated (as shown by the red dots in Fig. 2A). Upon applying a large positive gate bias, the whole channel becomes heavily protonated; thus, $α$-In$_2$Se$_3$H$_{x1}$ transforms into $α$-In$_2$Se$_3$H$_y$ (y>x$_1$). Subsequently, as the gate bias is successively reduced to 0 and then to a negative value, the number of protons across the channel eventually decreases to ~0 because of the electric-field-driven deprotonation (*24*). In this scenario, protons are gradually extracted from the channel, leading to a decreased proton population but still with a concentration gradient, decreasing from the channel's bottom surface to the top. Furthermore, in response to proton removal, the proton-sufficient $α$-In$_2$Se$_3$H$_y$ phase returns to proton-moderate $α$-In$_2$Se$_3$H$_{x2}$ at a gate bias of 0 V or proton-deficient $α$-In$_2$Se$_3$H$_i$ at a negative bias. It is noted that, in such a changing trend of gate biases, $α$-In$_2$Se$_3$H$_{x1}$ and α-In$_2$Se$_3$H$_{x2}$ have different protonation levels. In short, a large negative bias allows protons to be progressively driven out of the channel, resulting in the recovery of less protonic states (i.e., proton-deficient $α$-In$_2$Se$_3$H$_i$).

We used electrical hysteresis loops and *in-situ* PFM to examine proton-induced ferroelectric phase transitions in single-domain channels. As per $I_D$-$V_G$ transfer curves, our fabricated protonic transistors can be phenomenologically classified into two groups: the first is shown in Fig. 2B (Dev. 1), whereas the second is displayed in Fig. 4 (Dev. 2). To avoid the influence of electric-field induced polarization switching in these channels, a small voltage (~ 0.5 V) was used to read drain-source currents and monitor phase transitions. In the case of Dev. 1, there are gaps or spaces (see inset of Fig. 2B) introduced by the rough silica surface (fig. S6) and sample transfer at the sample/silica interface, which greatly restricts the efficiency of proton insertion at a given gate bias. Therefore, a gradual current change and corresponding small slope (rather than an abrupt change and large slope) in the $I_D$-$V_G$ transfer curves are observed (Fig. 2B). Remarkably, when increasing gate biases, the drain-source current increases, which is attributed to two coupling effects: electrochemical doping from proton injection (supported by the band structure calculations in fig. S7), and protonated phase changes. The predominant reason is not likely associated with electrostatic electron injection (nonvolatile characteristics in contrast with our volatile results; see Fig. 3 for the evidence), consistent with previous works (*24-26, 33*). As systematically illustrated in Fig. 2A, the hysteresis of Fig. 2B should be attributed to different protonation levels across the channel. For example, a 0-V gate bias acquired from either raising gate biases or lowing gate biases can lead to new disparate protonated phases. It is worth mentioning that $V_G \leq 3V$ in Fig. 2B cannot result in the dipole flipping of all $\alpha$-$In_2Se_3$ protonic phases, which can be evidenced by the closed loops in Fig. 5A.

With small bias (0.5 V) sweeping, a hysteresis loop can surprisingly appear (Fig. 2B), indicating the existence of a protonic phase in the channel. As the sweeping biases are progressively raised, the hysteresis becomes larger, which suggests that more protons move across the channel. We note that protonic hysteresis contrasts with that of electric-field-tuned ferroelectric devices (*34, 35*), which are always accompanied by an opening process (the hysteresis is only observed beyond certain threshold voltages). Additionally, it is found that sweeping speed has a small influence on the hysteresis (fig. S8).

To elucidate the phase transition with respect to hysteresis currents, we performed *in-situ* PFM mapping over a full sweeping cycle from -3 V to +3V and then back to -3 V. PFM is a reliable and powerful tool to monitor phase transitions via amplitude mapping. The brightness of PFM amplitude signal can be used to indicate the protonation levels of the channel because the protonic compound $\alpha$-$In_2Se_3$ has different electromechanical responses depending on proton concentrations. As shown in Fig. 2C and fig. S9A, there are uniform PFM responses across the channel, manifesting the presence of a single ferroelectric domain. Under different gate biases, the channels (marked by dashed black lines) display a spectrum of PFM amplitudes, which range from the response brighter to darker than the substrate. This is strong evidence for gate-bias-dependent protonation that typically proceeds from the channel bottom surface to the top. Meanwhile, identical PFM images, particularly in the case of 2 V and 3 V (or -2 V and -3 V) in Fig. 2C, demonstrate a trend of saturated protonation levels. Overall, a large gate bias ensures a high proton concentration across the channel, which facilitates heavy protonation; however, a small gate bias leads to light protonation. Furthermore, as shown in fig. S9B, we do not observe proton-induced sample height variations even at the gate biases of +3 and -3 V. By contrast, for gate biases of > 5 V, dramatic structural changes are accompanied by a large bump in the topographical profile (fig. S10). These observations can be explained by the presence of a gate-bias threshold for proton-triggered lattice expansion (*36, 37*). Importantly, gate-bias-dependent protonation can enable a multitude of intermediate ferroelectric phases, and thus facilitate the implementation of multilevel memory devices with massive storage capacity (*38*), which are promising for neuromorphic computing hardware.

**Volatile, gate-tuned protonic phases**

Having established the concept of protonic ferroelectric phase transitions using a single domain model, we herein look to the effect of protonation on multidomain phases. As shown in Fig. 3A, outer and inner patterns outlined by black dashed lines were written by applying -8 and +8 V onto PFM probes, respectively, leading to a change in PFM response. Such voltage poling enforces protonic $α$-$In_2Se_3H_x$ compound to orient outwards and inwards such that a multidomain phase is acquired. However, once a large negative gate bias (-3 V) was applied, proton extraction immediately takes place, leaving a barely protonated phase (i.e., proton-deficient $α$-$In_2Se_3H_i$). This can be inferred from distinct responses of PFM amplitude and phase in Fig. 3B. Note that, irrespective of domain orientations at a 0-V gate bias, proton extraction driven by -3 V gate bias makes the entire channel reverse to the proton-deficient phase, producing identical PFM responses for both inner and outer patterns. Fig. S11 shows the corresponding evolution of multidomain protonation under gate biases of < 3 V. This progressive evolution is in line with the behaviors in Fig. 2C. Unexpectedly, after the removal of gate bias, electric-field-based proton extraction rapidly terminates but proton reservoirs in the silica substrate continuously supplement the channels with protons, consequently reversing $α$-$In_2Se_3H_i$ to $α$-$In_2Se_3H_x$ ($x>i$) phase, as revealed by PFM mapping in Fig. 3C. This behavior verifies the volatile nature of ferroelectric phase transition induced by proton injection, distinguishable from electric-field induced nonvolatility (*39*). The volatility is coincident with that of proton-induced ferromagnetic-paramagnetic phase transition (*26*).

To give insight into the volatile PFM response, we obtained dynamic drain-source currents as a function of the application and termination of gate ($V_G$) or read ($V_{DS}$) biases. Overall, a negative 3-V gate bias can sharply reduce the drain-source current from initial states; after the removal of this gate bias, the current increases but remains much smaller than the initial value (Fig. 3D). However, once the read bias is turned off, the drain-source current coded by a -3 V gate bias is completely erased and returns to the initial value. These electrical properties are a remarkable signature of volatility for gate-tuned protonic phases, which coincides with our PFM results. Interestingly, the trend in the current change—a peak and saturation currents—upon the application of -3 V gate bias suggests that protonation can finally reach a saturated plateau across the channel (i.e., barely protonated). This is additional clear evidence for proton-induced phase transition as observed in $VO_2$ (ref. (*24, 33*)). After taking away the gate bias ($V_G = 0$), metastable protonic phases relax to an energetically stable state, i.e., $α$-$In_2Se_3H_x$ compound. Despite a small conductance change, a reversal of $α$-$In_2Se_3H_x$ can occur, demonstrating the sensitivity of ferroelectric phase transitions to proton concentration/gate biases. Similar results are also demonstrated for gate tuning by positive gate biases (fig. S12).

**Low switching voltages and polar switchability**

Next, we discuss the second type of protonic $α$-$In_2Se_3$ devices with a high proton injection efficiency (Dev. 2; Fig. 4). Compared to Dev. 1 in Fig. 2, the generated spaces or gaps at the interfaces are largely reduced because channel samples in this device are almost in complete contact with the silica substrate. Possibly because of this ideal interface, abrupt changes in transfer curves are obtained as shown in Fig. 4A. For the transition from off-current to on-current, an ultralow voltage (< 0.4 V) is sufficient to provide the required energy for complete protonation across the channel. As the voltage sweeps back in the other direction and changes polarity, protons are dramatically driven out of the channel, creating a proton-deficient $\alpha$-In$_2$Se$_3$H$_i$ phase.

To have an in-depth understanding of electrical hysteresis loops in Fig. 4A, PFM measurement was employed to analyze the change in material composition. Albeit using a small gate bias (0.5 V), PFM amplitude discrepancies in Fig. 4B clearly affirm proton-induced phase transition. Furthermore, similar electrical and PFM behaviors were also detected in a separate transistor (fig. S13), which shares the same channel flake with the device used to produce the data in Fig. 4 but without bottom gate (BG) electrodes. Such a control experiment confirms the facts of proton diffusion within channel materials and proton-induced hysteresis, which rule out the electric-field-induced switching. Note that the ultralow operation voltage (< 0.4 V) provides an encouraging possibility of developing low-power information storage devices. Unlike ferroelectric-polarization-based memory devices (see Table 1), our protonic ferroelectric transistors have competitive advantages in terms of the gate switching voltage.

Finally, as indicated by the blue and red curves in Fig. 5, we utilized electrical hysteresis loops to uncover reversible polarization switching for heavily and barely protonated phases. Under low drain-source voltage sweeping, there is no obvious hysteresis in both $\alpha$-In$_2$Se$_3$H$_i$ and $\alpha$-In$_2$Se$_3$H$_y$ phases because the field magnitude is lesser than their required coercive fields. However, when drain-source voltages are increased to 5 V, an expanded hysteresis is observed with a high on/off ratio. These comparable results, in addition to theoretical predictions in Fig. 1 and multiple domain patterns in Fig. 3A, strongly demonstrate that protonated $\alpha$-In$_2$Se$_3$ phases maintain switchable dipoles. The newly discovered polar nature is nontrivial due to the incorporation of two mutually exclusive phenomena, i.e., ferroelectricity and ion conduction, particularly at high proton concentrations.

**Conclusion**

Using van der Waals ferroelectric $\alpha$-In$_2$Se$_3$ as a model system, we have created a multitude of metastable ferroelectric phases and demonstrated their reversible transitions by gate-bias-induced proton injection. We find that the produced ferroelectric phases are volatile and polar, which can be attributed to hydrogen-stabilized $\alpha$-In$_2$Se$_3$ phases with switchable dipoles. Moreover, we also demonstrate the ultralow voltage switching of ferroelectric phases (< 0.4 V).

The ability to intercalate ions into van der Waals ferroelectrics opens up a new route for exploring collective physical interactions, such as the effects of charge, lattice, ion, and dimensionality. Although our calculation has predicted that the inserted protons form covalent bonds with Se atoms, the possibility of proton migration in the channels via the continuous bond-breaking and formation process remains unknown. Along this line, new opportunities may emerge, such as the gate-tunable coupling of ion migration and electric dipoles, as well as their thickness-dependent effect. Furthermore, these inserted ions are envisioned to provide additional energy, during ferroelectric switching, to stabilize certain hidden phases involved with covalent-bond breaking, which are barely accessible in equilibrium phase states. In addition, the insertion of ions into van der Waals ferroelectrics has the potential to achieve numerous bits of information storage (fig. S14) in a single transistor cell, by finely tuning the quantity of created new phases. This prospect awaits further confirmation.

## MATERIALS AND METHODS

### Protonic transistor fabrication

First, a 10-nm-thick Pt gate electrode (50 μm × 50 μm) was fabricated on a $SiO_2$/Si substrate via electron beam lithography and metal sputtering. Then a 10 nm-thick $Al_2O_3$ layer was deposited over the entire substrate using atomic layer deposition to reduce gate leakage current. Afterward, a 190-nm-thick porous silica layer, which acts as a proton reservoir, was coated using a previously reported method (*24*). Tetraethyl orthosilicate, ethanol, deionized water, and phosphoric acid (85 wt%) were mixed in a molar ratio of 1:18:5.55:0.02, respectively, and stirred in a sealed bottle for 1 h. After annealing in an oven at 60 °C for 2 h, the solution was spin-coated on $Al_2O_3$/Pt/$SiO_2$/Si substrates at 3000 rpm for 1 min, followed by annealing on a hot plate at 120 °C for 30 min. Subsequently, α-$In_2Se_3$ flakes exfoliated from the parent crystal via the scotch tape method were transferred onto PDMS (Polydimethylsiloxane) stamps. α-$In_2Se_3$ flakes with appropriate shapes and thicknesses were identified using optical microscopy and then transferred onto porous silica layers using a micro-manipulator (Metatest E1-M), just on top of the Pt gate electrode. After electron beam lithography patterning and electron beam evaporation, Ti/Au (10 nm/70 nm) electrodes were deposited on the α-$In_2Se_3$ flakes to produce protonic transistors.

### Electrical, Raman, and PFM measurements

All electrical measurements were performed at room temperature by using a Keithley 4200-SCS semiconductor characterization system equipped with SMU modules. A probe station (Everbeing EB-6) was used to support these devices and connect microelectrodes with the semiconductor characterization system.

Raman spectra were collected using WITech alpha300 with a 532 nm laser and 1800 g/mm grating. The laser power density was set to ~ 0.3 mW to avoid damaging α-$In_2Se_3$ flakes. A 50× objective lens was used to monitor the Raman signals.

PFM was conducted by using Asylum MFP-3D with a probe constant of 2.8 N/m. An AC bias of 0.5 V and dual AC Resonance Tracking (dart) mode were adopted to acquire amplified out-of-plane piezoelectric responses. During the measurements, source terminals were always grounded to reduce charge accumulation across sample surfaces. A Keithley 2636B voltage source was employed to apply a DC gate bias during PFM measurements.

### First-Principles Calculations

First-principles calculations for structural relaxation and ground state electronic structure were performed using density-functional theory (*40, 41*) as implemented in the Vienna Ab initio Simulation Package (VASP) (*42, 43*) with the projector-augmented wave method for treating core electrons. The maximal residual forces for structural relaxation were less than 0.02 eV Å$^{-1}$, and the convergence criteria for electronic relaxation was $10^{-6}$ eV. We employed the Perdew-Burke-Ernzerhof (PBE) (*44*) form of exchange-correlation functional, a plane-wave basis with an energy cutoff of 400 eV, and a Monkhorst-Pack k-point sampling of 12×12×1 for the Brillouin zone integration. The climbing image nudged elastic band method (*45*) was adopted for searching the minimum energy pathways based on the interatomic forces and total energy acquired from DFT calculations.

## Acknowledgments

This research was supported by the ZJU-Hangzhou Global Scientific and Technological Innovation Center with a startup funding (02170000-K02013012) and the King Abdullah University of Science and Technology Office of Sponsored Research (OSR) under award number: ORA-CRG8-2019-4081 and ORA-CRG10-2021-4665. K. L. acknowledges support from the SRC/NIST nCORE SMART center. W. H. would like to acknowledge the support by the National Science Foundation of China (Grant Nos. 61974147).

## Author contribution

X. H. and F. X. together conceived and designed the experiments. X. H. fabricated the nanodevices and measured their electronic properties. F. X. performed the PFM characterizations. X. H and Y. M carried out the Raman spectra measurements. C. Z. provided the silica wafers with marks. H. W and A. F. performed the first-principles calculations. F. X., X. H., H. W., and Y. M. analyzed the data. W. H., Y. X., B. Y., and K. L. helped improve the paper. F. X and X. H. wrote the paper. X. Z. and F. X. supervised the project.

## Competing interests

The authors declare no competing interests.

## Data availability

The data supporting the findings of this study are available within the article and its supplementary files and available from the authors upon reasonable request.

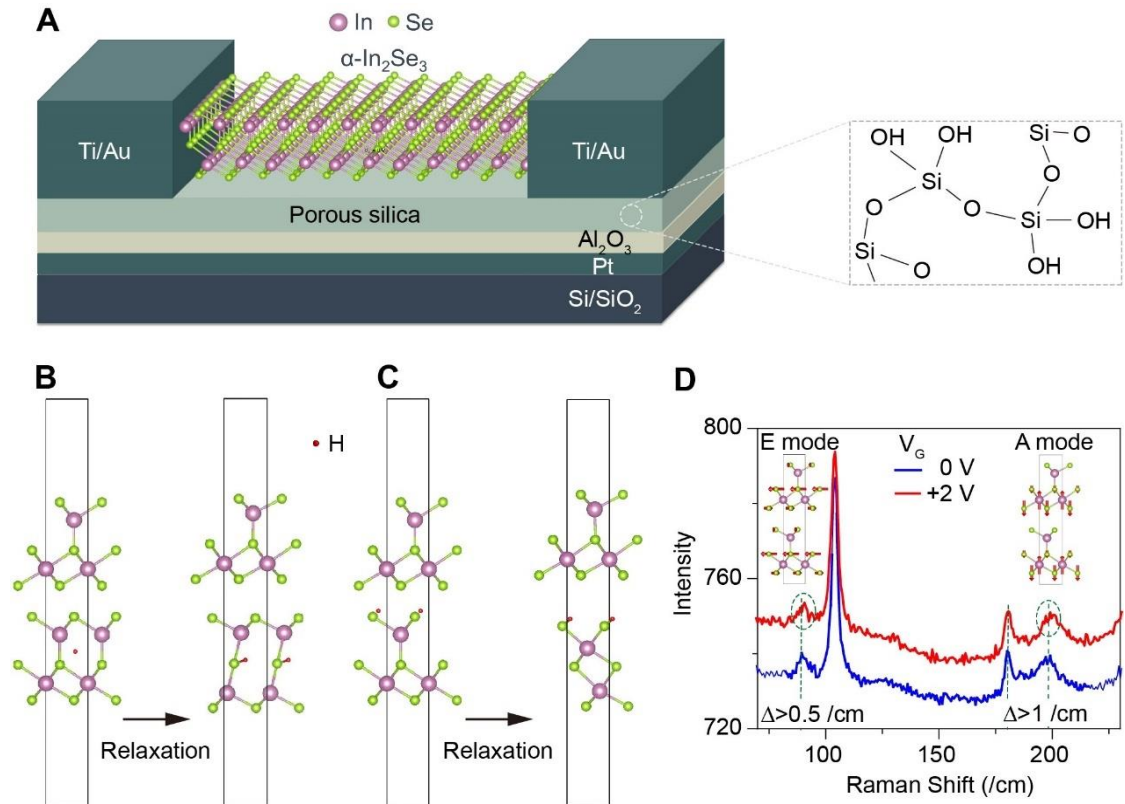

**Fig. 1. Device structure and working principle.** (**A**) Schematics of our three-terminal α-In$_2$Se$_3$ devices. Exfoliated α-In$_2$Se$_3$ flakes were placed on stacked heterostructure gates, which are composed of porous silica, Al$_2$O$_3$, and Pt. Right panel: the chemical structure of porous silica. First-principles structural relaxation of protonic α-In$_2$Se$_3$: H atom (**B**) at the intralayer and (**C**) at the van der Waals gap. (**D**) Raman spectra under 0-V and +2-V gate biases. Dashed circles highlight a blue shift of Raman peaks, which are enhanced by the proton injection across the van der Waals gaps. For E and A$_1$ modes, schematics of calculated in-plane and out-of-plane phonon vibrations are shown in the inset, respectively.

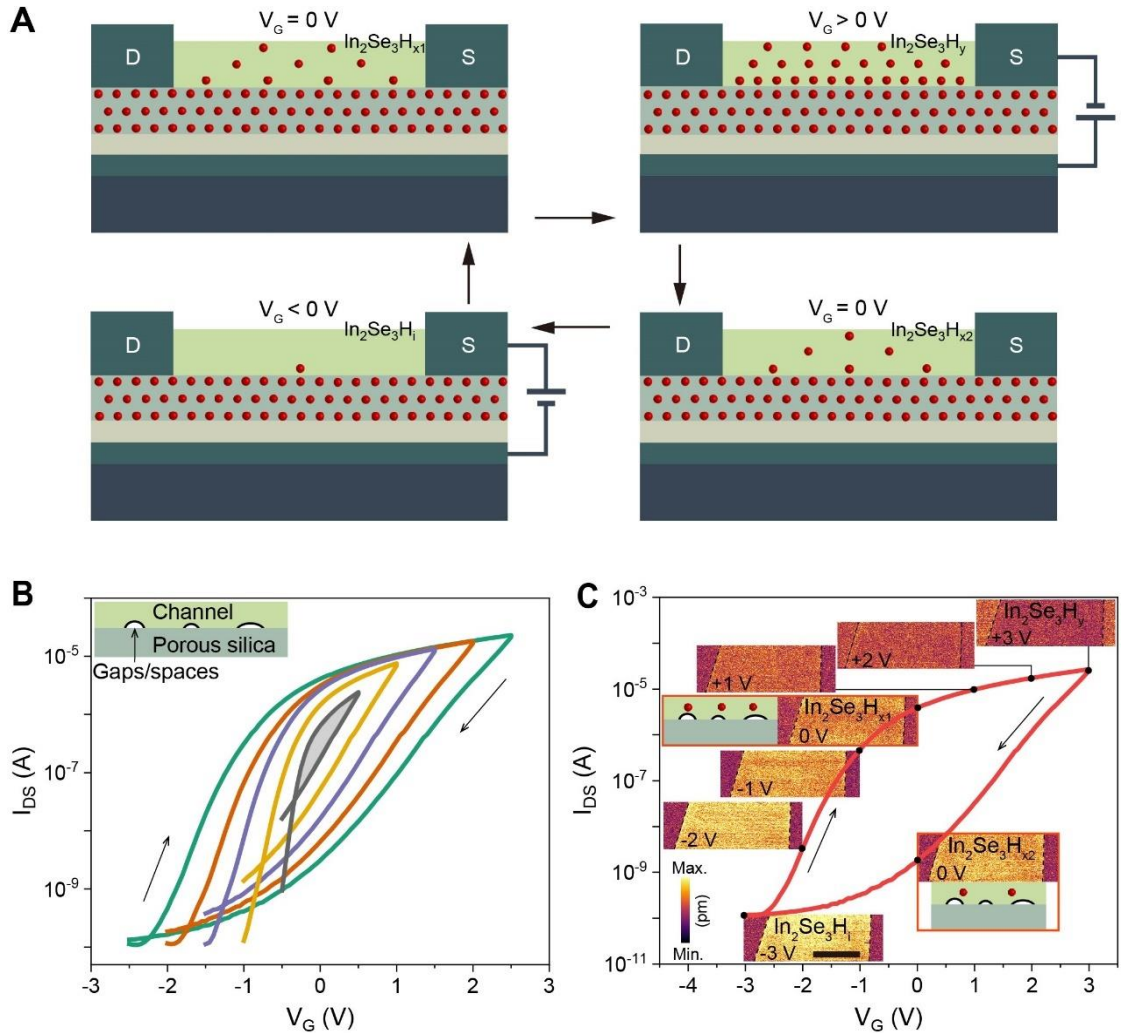

**Fig. 2. Reversible switching of multiple ferroelectric phases by a protonic gate for Dev. 1.** (**A**) Schematic working principle of ferroelectric phase transitions under different gate biases. Red dots in the porous $SiO_2$ layer represent protons. At a 0-V gate bias, thermal activation leads to an inherent proton penetration into the channel, forming the $α$-$In_2Se_3H_x$ compound. When applying a large, positive gate bias, many protons (red dots) are pushed into the channel, achieving heavy protonation across the channel (i.e., $α$-$In_2Se_3H_y$ phase). Afterward, when the gate bias is returned to 0 V, protonation still exists but to a lesser content than that in the top-left panel. As the gate bias becomes negative, protons can be driven out of the channel, and the composition returns to one close to the primary $α$-$In_2Se_3$ phase (i.e., $α$-$In_2Se_3H_i$). (**B**) Electrical hysteresis with respect to maximum gate biases. The first maximum voltage starts at 0.5 V and then increases in a 0.5-V step. The inset shows interfacial contacts between porous silica and the channel. (**C**) Phase evolution as a function of gate-bias sweeping. PFM amplitude mappings, at the gate biases of -2, -1, 0, +1, +2, +3, 0, and -3 V, demonstrate the evolution of ferroelectric phases (rather than polarization switching) between proton-sufficient $α$-$In_2Se_3H_y$ and proton-deficient $α$-$In_2Se_3H_i$. For $α$-$In_2Se_3H_x$ compounds, schematic proton distributions are shown to differentiate their discrepancies. The $α$-$In_2Se_3$ channel is highlighted by dashed black lines. Note that $V_G ≤ 3V$ here is insufficient to reverse the polarization of protonated $α$-$In_2Se_3$. Scale bar: 2 μm.

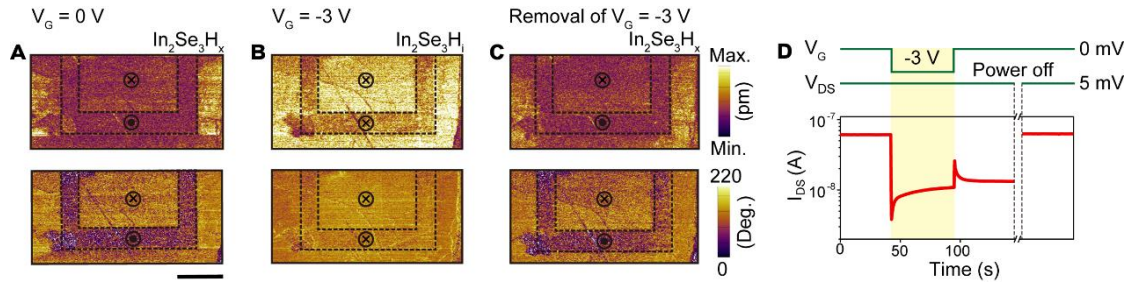

**Fig. 3. Volatility of protonated ferroelectric phases.** Patterned ferroelectric domains under different gate biases: (**A**) 0 V, (**B**) -3 V, and (**C**) after the removal of -3 V. At a 0-V gate bias, outer and inner boxes in (a) were written by PFM probes with DC voltages of -8 and +8 V, respectively. "⊙" and "⊕" indicate polarization pointing upward and downward, respectively. The top panels in **A-C** are PFM amplitude mappings while the bottom panels are PFM phase mappings. (**D**) Dynamic current change upon applying different $V_G$ and $V_{DS}$. Although the application of $V_G$ can switch the channel current to a low level, this switching effect can be erased by removing source-drain voltage, suggesting the volatile nature of protonic modulation. Scale bar: 1 μm.

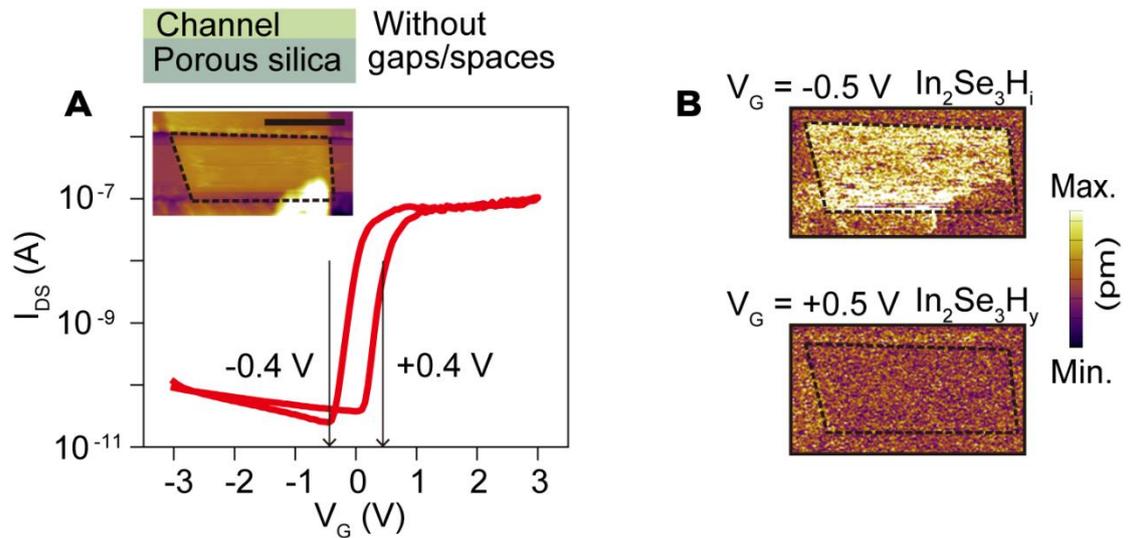

**Fig. 4. Ferroelectric phase transitions with low operation voltages (Dev. 2).** (**A**) Electrical hysteresis loops for proton-gated $\alpha$-In$_2$Se$_3$ devices with ideal interfaces (i.e., no gaps/spaces between the channel and the porous silica). The inset shows an AFM topographic image of a typical device, in which dashed lines highlight the device channels. Scale bar: 1 μm. (**B**) PFM amplitude mappings of the channel area (i.e., the inset of panel **A**) with the application of different gate biases: -0.5 V and +0.5 V.

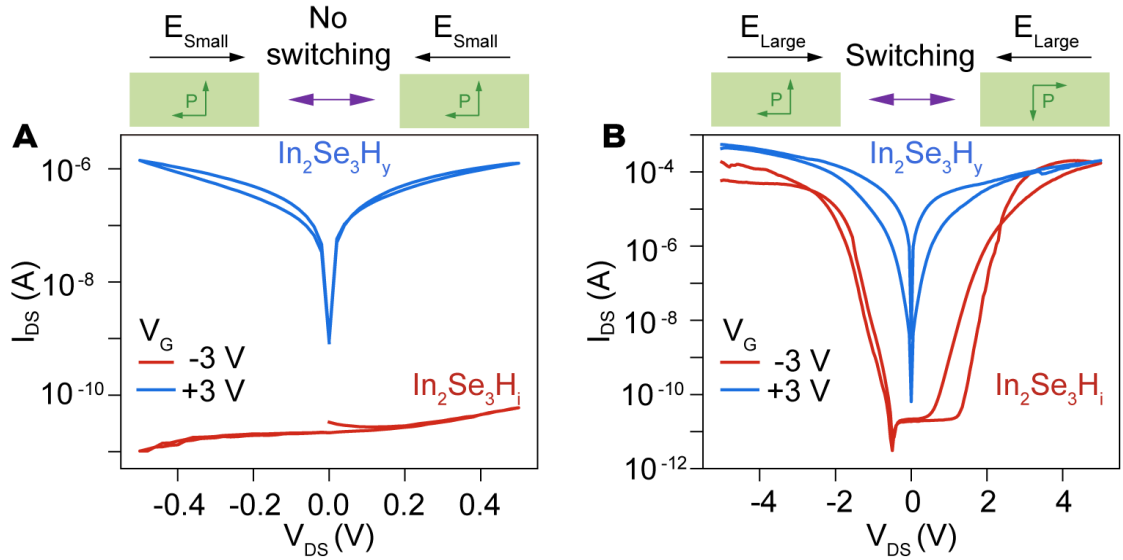

**Fig. 5. Evidence for the switchability of proton-deficient and proton-sufficient phases.** Electrical hysteresis loops under different gates and read biases: drain-source voltage sweeps over (**A**) ±0.5 V and (**B**) ±5 V. Schematics on top of (**A**) and (**B**) indicate whether the switching of protonated α-In$_2$Se$_3$ phases occurs under small (E$_{Small}$) and large (E$_{Large}$) in-plane electric fields. Black arrows dictate the directions of sweeping electric fields, whereas green orthogonal arrows represent the interlocked in-plane and out-of-plane polarizations in α-In$_2$Se$_3$H compounds.

**Table 1. Comparison of gate switching voltages**

| Device structures | Gate switching voltage | Dielectric thickness | Mechanism |
|---|---|---|---|
| AlScN/MoS$_2$ FE-FET (46) | -40 V to +20 V | 100 nm | Electric field-induced polarization switching |
| CIPS/h-BN/InSe FE-FET (47) | -4 V to +1 V | 10 nm CIPS | |
| Gr/h-BN/PMN-PT FE-FET (48) | -50 V to +50 V | 100 μm PMN-PT | |
| MoS$_2$/CIPS FE-FET (49) | -3 V to +3 V | 400 nm CIPS | |
| In$_2$Se$_3$ FS-FET (50) | -40 to +5 V | 90 nm SiO$_2$ | |
| In$_2$Se$_3$ FS-FET This work | -0.4 V | 10nm Al$_2$O$_3$/190 nm porous silica | Proton-induced ferroelectric phase transition |

FE-FET: ferroelectric field-effect transistors; FS-FET: ferroelectric semiconductor field-effect transistors; CIPS: CuInP$_2$S$_6$; Gr: graphene; and PMN-PT: Pb(Mg$_{1/3}$Nb$_{2/3}$)O$_3$–PbTiO$_3$.

# Supplementary Materials

**Note S1. Ferroelectric α-In$_2$Se$_3$**

α-In$_2$Se$_3$ is an emerging van der Waals (vdW) ferroelectric semiconductor with structural distortion from middle Se-In-Se atomic layers (*51*). Its ferroelectric polarization can persist even down to a monolayer thickness (*52*) because spontaneously interlocked in-plane (IP) and out-of-plane (OOP) dipoles resist the perturbation from the depolarization field (*53*). This dipole locking mechanism ensures that the switching of OOP polarization can lead to IP polarization switching, and vice versa (*54*). Due to the attributes of robust ferroelectricity at the atomic limit, a moderate direct bandgap (~1.4 eV–1.6 eV) (*55*), and high Curie temperature (above 700 K) (*53*), two-dimensional α-In$_2$Se$_3$ has been used in a broad range of memory device applications, including ferroelectric field-effect transistors (*50*), ferroelectric semiconductor junctions (*56*), and ferroelectric optoelectronic memories (*57*). Furthermore, these devices have been implemented with high performance (large on/off ratio and multilevel switching) for next-generation computing paradigms, such as neuromorphic computing and in-memory computing (*34, 58*). It is worth mentioning that α-In$_2$Se$_3$ has been successfully grown by various methods, which demonstrates its potential for practical applications (*51*).

**Note S2. Mechanism of proton generation in porous silica**

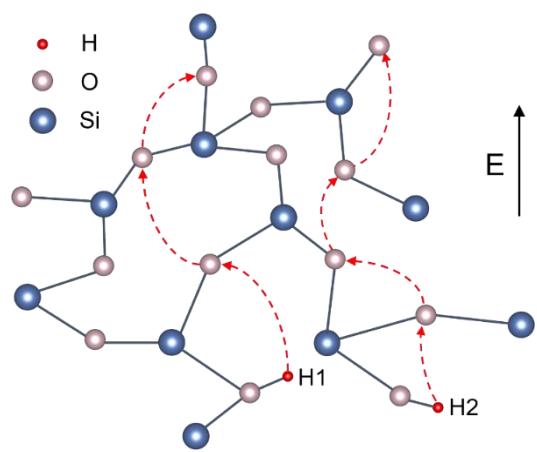

**Fig. S1. Diagram for proton hopping in porous silica under an external electric field.**

As shown in fig. S1, porous silica, which is commonly fabricated by low-temperature methods (e.g., sol-gel methods), is amorphous with many dangling bonds. Because of the phosphoric acid introduced in the material fabrication process, plenty of protons ($H^+$) exist in the form of hydroxyl groups (-OH) inside silica. As indicated by H1 and H2 in fig. S1, these protons are bonded with oxygens in the form of Si-OH-Si and Si-OH, respectively (*24, 59, 60*). The bonds between oxygen and hydrogen are not stable and can be broken easily (*60*). Therefore, protons are mobile and can hop to neighboring oxygen atoms (*60, 61*), forming new -OH. In other words, protons can migrate across the Si-O networks of porous silica by breaking and forming -OH continuously. These mobile protons can either diffuse from regions with a high density of protons to those with a low density, or drift along the direction of an external electric field (fig. S1) (*24, 60*); both of these mechanisms are similar to the migration behavior of positive holes in semiconductors.

The other pathway for producing protons is the electrolysis of water. There is remaining water in the porous silica since the temperature of the sol-gel method is not high enough to desorb all water molecules from the film (*59*). In addition, numerous nanopores inside porous silica can easily absorb water vapor from the air via the capillary effect (*24*). When the porous silica is connected to a circuit, the interior water can be electrolyzed into $H^+$ and $O_2$ in the anode by applying a voltage above 1.23 V (the electrode potential for water electrolysis) (*24*). The reaction in the anode is as follows:
$$2H_2O \rightarrow O_2 + 4H^+ + 4e^- \quad (1)$$
Therefore, this electrolysis process can provide abundant protons continually for channel protonation (*24*). Due to the unique properties of protons inside porous silica, the capacitance per unit area is about two orders of magnitude higher than that of the typical $SiO_2$ (*62*), providing an efficient method to modulate the channel conductance.

**Note S3. Mechanism of protonation and deprotonation in α-In$_2$Se$_3$**

When porous silica and α-In$_2$Se$_3$ form a transistor-like three-terminal device, the porous silica acts as a solid electrolyte. As a positive gate bias is applied to the porous silica, protons drift from silica to α-In$_2$Se$_3$, and the same number of electrons flow into α-In$_2$Se$_3$ from the external circuit simultaneously to maintain charge neutrality (*24*), which resembles the charging process of a Li-ion battery (*63*). These protons intercalate into the α-In$_2$Se$_3$ flakes and form α-In$_2$Se$_3$H$_x$ compounds together with injected electrons, consequently increasing the channel conductivity (*24*):

$$\alpha\text{-In}_2\text{Se}_3 + x\text{H}^+ + x\text{e}^- \rightarrow \alpha\text{-In}_2\text{Se}_3\text{H}_x \quad (2)$$

This process is termed protonation. In contrast, when the gate bias is decreased, protons begin to de-intercalate from α-In$_2$Se$_3$, and electrons flow back to the external circuit simultaneously, decreasing the conductivity of the α-In$_2$Se$_3$ channel (*24*):

$$\alpha\text{-In}_2\text{Se}_3\text{H}_x \rightarrow \alpha\text{-In}_2\text{Se}_3 + x\text{H}^+ + x\text{e}^- \quad (3)$$

This process is termed deprotonation. When the gate bias becomes negative, the deprotonation process accelerates until all protons are driven out of the α-In$_2$Se$_3$ channel. It is worth noting that the energies required for intercalation (protonation) and deintercalation (deprotonation) are different (*64*). Thus, the protonation levels are different at the same bias in the forward (protonation) and backward (deprotonation) sweeping directions of the gate bias, creating two different protonation phases. As a result, the electron densities (conductivity) determined by the protonation level are different at the same gate biases, leading to hysteresis in the transfer curve. Moreover, like the relatively slow movement of ions in ionic liquid, protons may also have a delayed response to the electric field, which can contribute to the hysteresis loop (*65*).

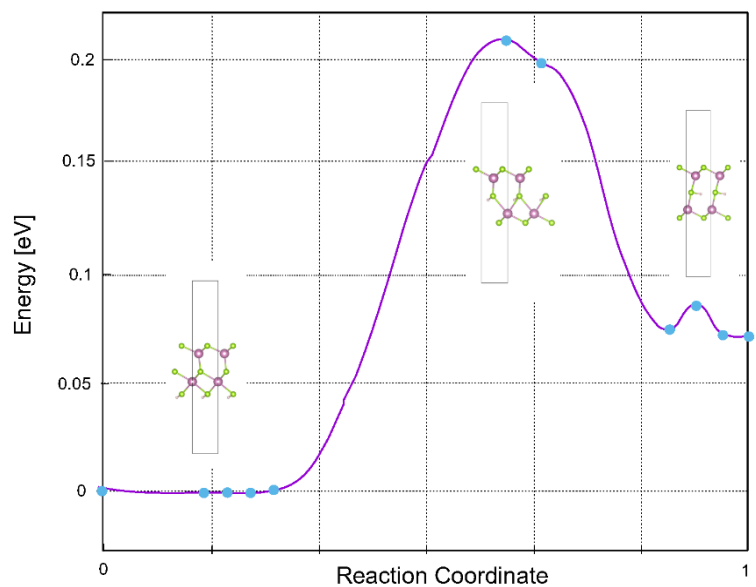

**Fig. S2.** Minimum energy pathways (MEP) of the protonation process in monolayer $\alpha$-In$_2$Se$_3$ calculated using the climbing image nudged elastic band (NEB) method for the band structures of monolayer $\alpha$-In$_2$Se$_3$ and $\alpha$-In$_2$Se$_3$H.

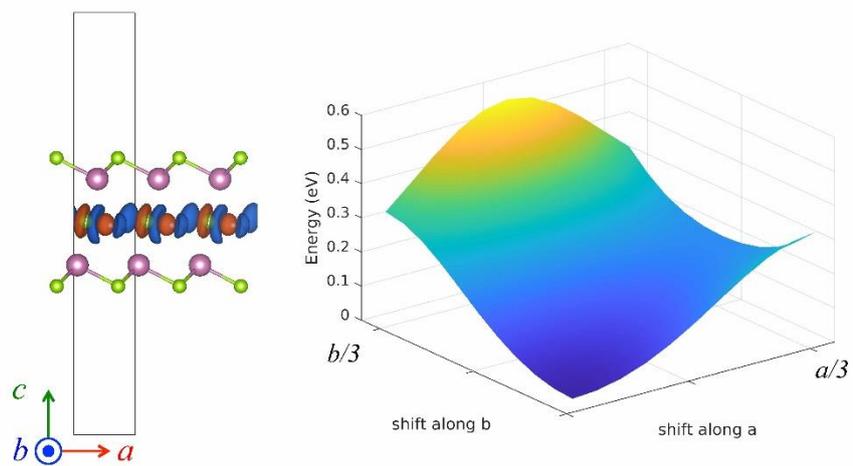

**Fig. S3.** Charge difference calculated by subtracting the valence electron density of individual H atoms in monolayer $\alpha$-In$_2$Se$_3$H (left panel) and generalized stacking fault energy surface in monolayer $\alpha$-In$_2$Se$_3$ (right panel).

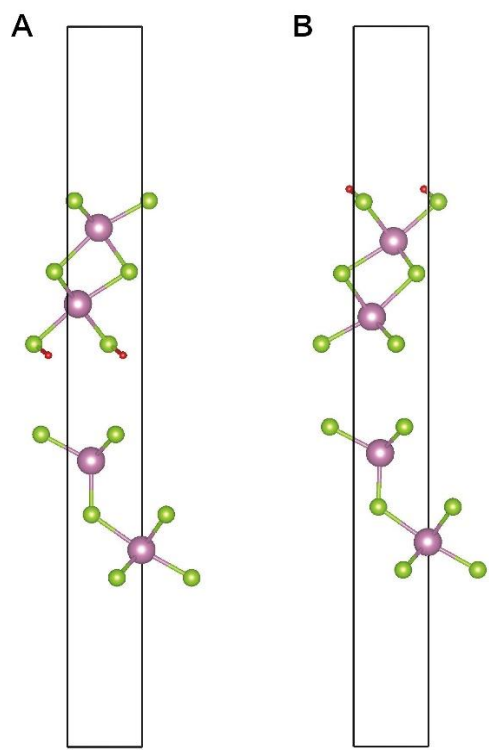

**Fig. S4. Other possible scenarios of hydrogen positions in *α*-In$_2$Se$_3$H compounds.** Calculated structures for relaxed protonic *α*-In$_2$Se$_3$ bilayers with (**A**) hydrogens fixed at the van der Waals gap but closer to the top quintuple-atom-layer, and (**B**) hydrogens at the top surface.

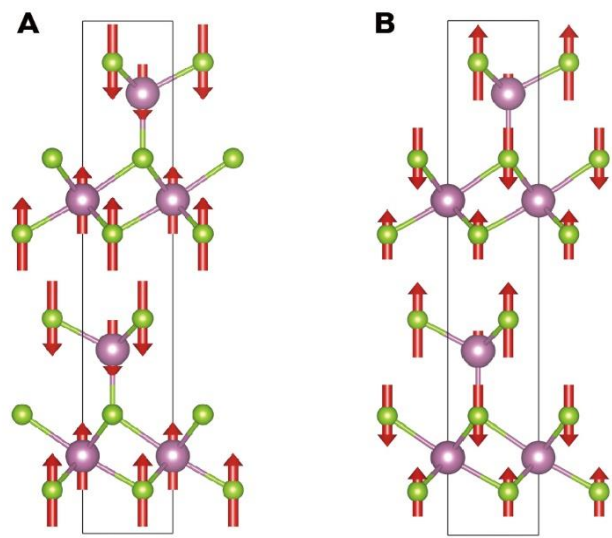

**Fig. S5.** Schematics of Raman phonon vibrations related to strong intra-layer breathing for the peaks located at (**A**) 104 and (**B**) 180 cm$^{-1}$.

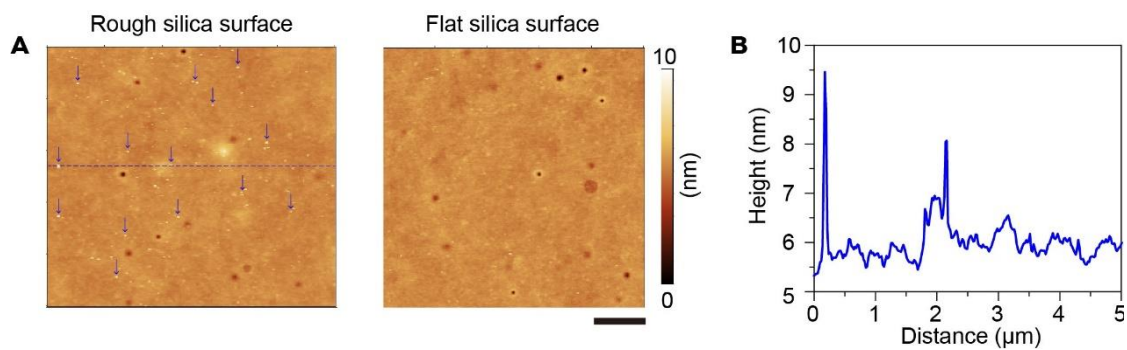

**Fig. S6. Roughness of different silica surfaces.** (**A**) PFM topographic images for the devices with low efficiency of proton injection (left panel), and high efficiency of proton injection (right panel). The left PFM image clearly shows a larger roughness, which is likely caused by crystalline silica nanoparticles (typical bumps are highlighted by blue arrows). Scale bar: 1 μm. (**B**) Height profile taken along the dashed blue line in (**A**) showing that the diameters of the nanoparticles were approximately 0.5–4 nm.

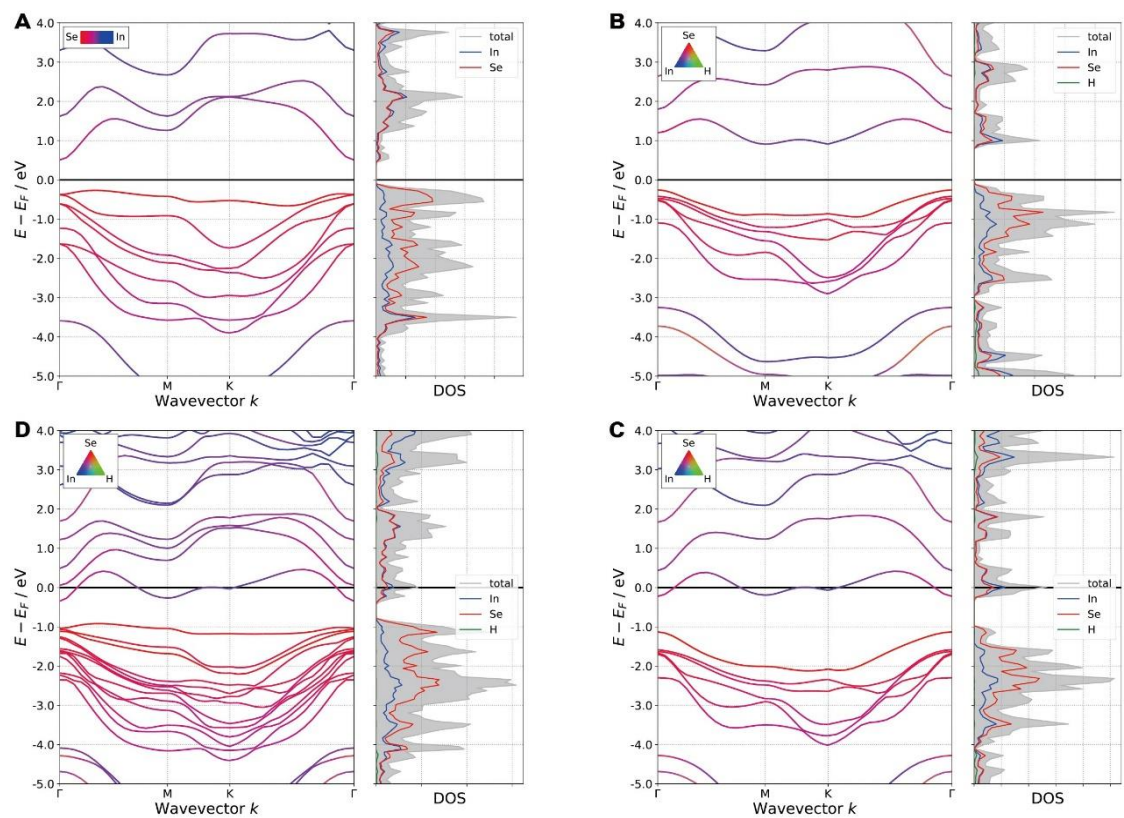

**Fig. S7.** Band structures for (**A**) monolayer $In_2Se_3$, (**B**) $In_2Se_3H^+$, (**C**) $In_2Se_3H$, and (**D**) bilayer $In_2Se_3H$. The upward shift of the Fermi level contributes to the conductivity increase observed in Fig. 2.

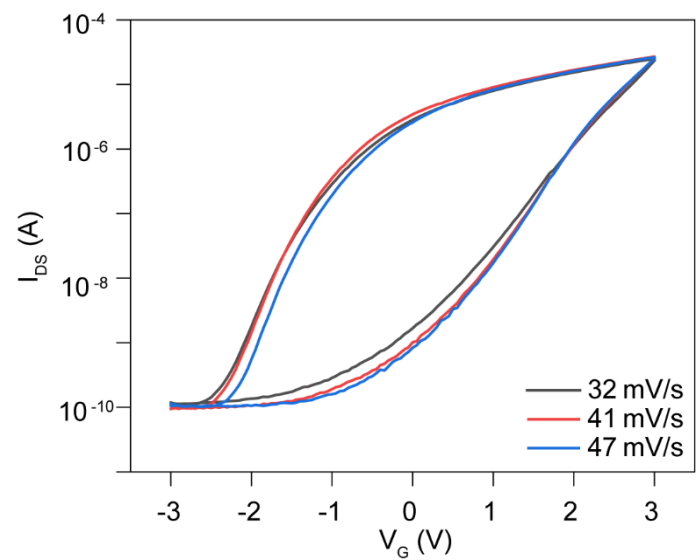

**Fig. S8.** Electrical hysteresis loops as a function of sweeping rates. The sweeping rate has little effect on the hysteresis.

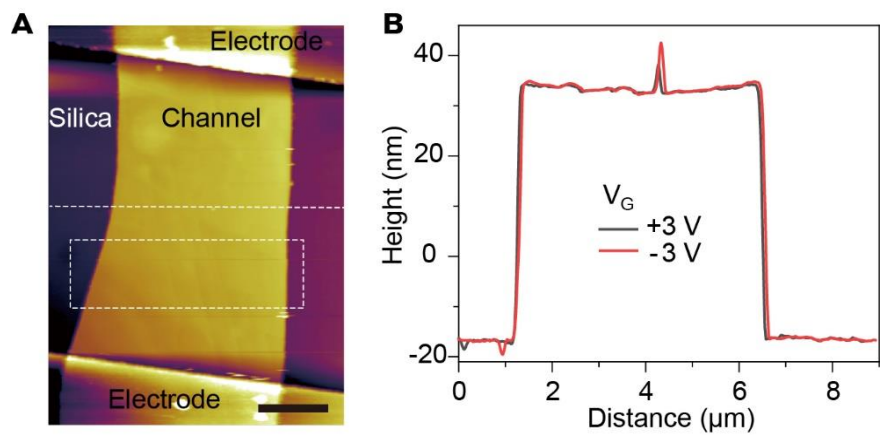

**Fig. S9.** (**A**) PFM topographic image for the device in Fig. 2. The set of PFM amplitude mappings in Fig. 2c was collected from the area indicated by the dashed white rectangle. Scale bar: 2 μm. (**B**) Height profiles as a function of gate bias. The data were measured along the dashed white line in (**A**). In the cases of protonation (+3 V) and deprotonation (-3 V), there is no obvious height change, suggesting that the threshold of proton-induced structural expansion exceeds 3 V.

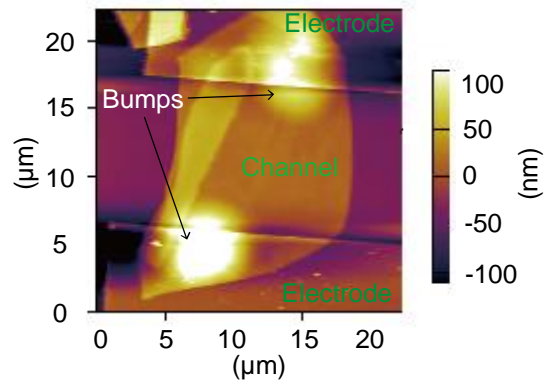

**Fig. S10. AFM topography of a typical device.** As the gate bias is increased to over 5 V, excessive protons flow into and fully protonate the channel. Consequently, structural expansion occurs due to proton-induced lattice variation, and a few marked bumps are observed.

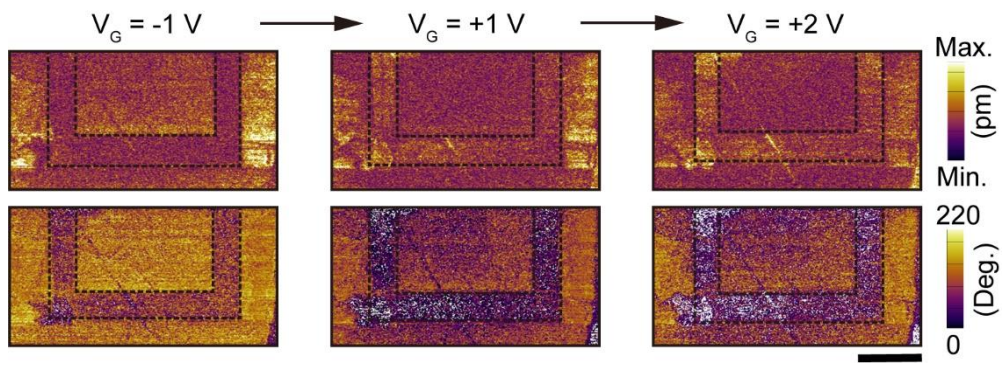

**Fig. S11.** Evolution of the PFM amplitude (top row) and phase mappings (bottom row) with different gate biases for the device in Fig. 3. Scale bar: 1 μm.

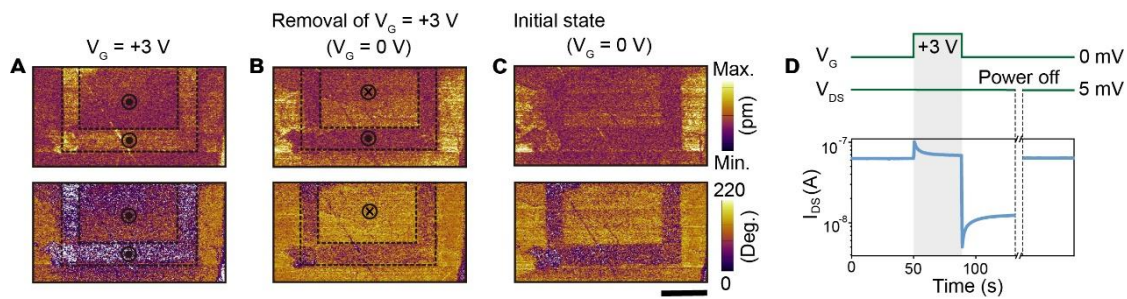

**Fig. S12.** (A-C) PFM amplitude (top) and corresponding PFM phase (bottom) of the sample in Fig. 3 at the protonic gate voltage of (**A**) +3 V, (**B**) 0 V (after removal of the +3 V), and (**C**) 0 V (the initial state). Scale bar: 1 μm. (**D**) Dynamic current changes upon the application of different $V_G$ and $V_{DS}$.

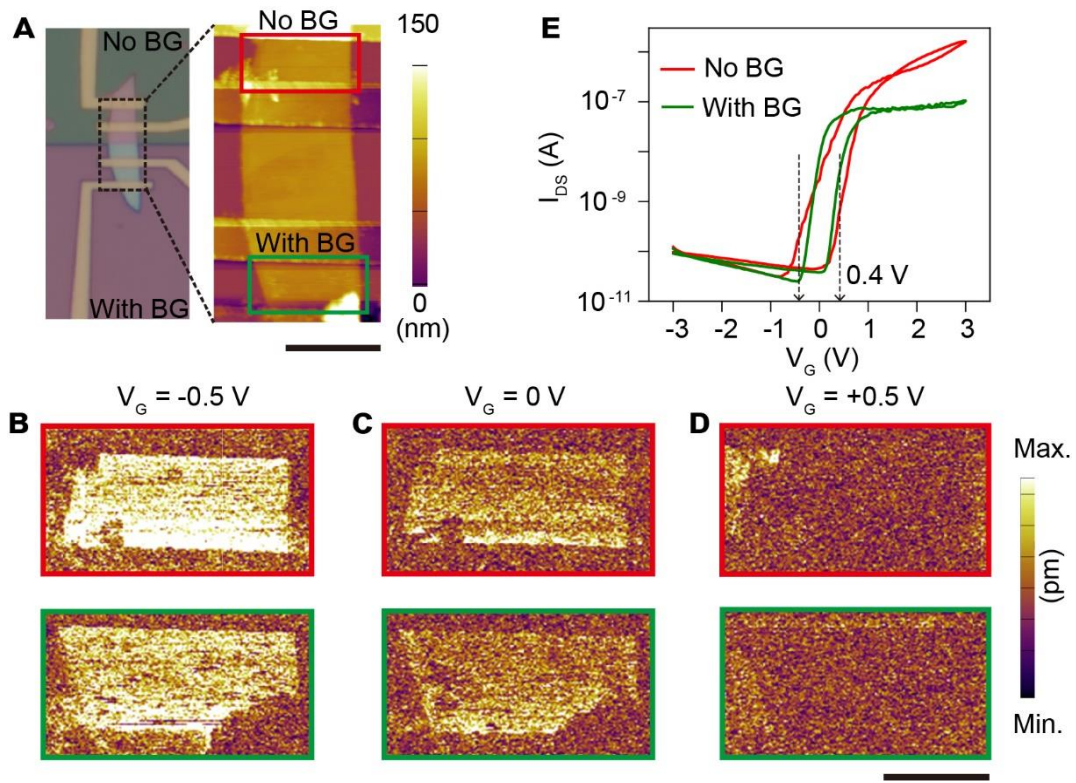

**Fig. S13. Control experiments.** (**A**) Optical (left) and enlarged AFM (right) images for a typical α-In$_2$Se$_3$ device. The lower two electrodes bridge a channel on the porous silica/Pt substrate (with bottom gate (BG)), forming a protonic transistor as described in the main text. In contrast, the upper two electrodes connect a channel just on a silica substrate, rather than silica/Pt (i.e., no BG). Scale bar: 2 μm. (**B-D**) PFM amplitudes for the upper (red) and lower (green) channels under various protonic gate voltages. Scale bar: 1μm. Note that, for comparison, we also plotted the data in Fig. 4. (**E**) Electrical hysteresis loops for both upper and lower channel devices.

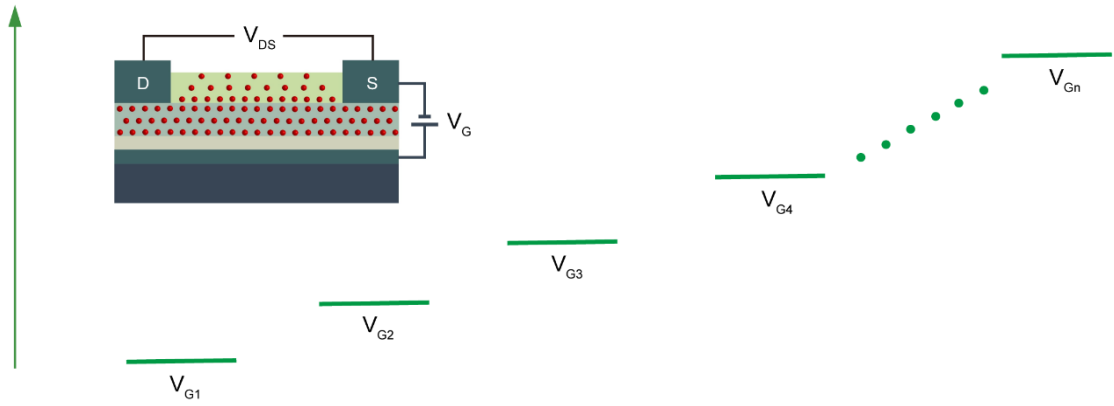

**Fig. S14. Concept of multilevel resistance switching in protonic ferroelectric transistors.** To successfully achieve a multitude of phases, DC gate biases ($V_G$) are always required to maintain the protonic states. Once a DC gate bias is set on the device, a drain-source voltage ($V_{DS}$) can be used to switch channel domains and store information.